\documentclass[12pt,oneside]{article}
\usepackage{amsfonts,amssymb,graphicx}
\def\be{\begin{equation}}
\def\ee{\end{equation}}
\def\bea{\begin{eqnarray}}
\def\eea{\end{eqnarray}}
\def\ba{\begin{eqnarray*}}
\def\ea{\end{eqnarray*}}

\def\<{\langle}
\def\>{\rangle}
\def\~{\tilde}
\def\s{\sigma}

\def\b{\beta}

\def\ds{\displaystyle}

\newcommand{\av}[1]{\mbox{{\rm Av}}\left[#1\right]}

\begin{document}
%
%
\begin{center}
\vspace{1truecm}
{\bf\sc\Large lack of monotonicity in spin glass\\ correlation functions}\\
\vspace{1cm}
{Pierluigi Contucci $^{\dagger}$,\quad Francesco Unguendoli $^{\ddagger}$, \quad Cecilia Vernia $^{\ddagger}$, }\\
\vspace{.5cm}
{\small $^\dagger$ Dipartimento di Matematica, Universit\`a di Bologna, {e-mail: {\em contucci@dm.unibo.it}}} \\
\vspace{.5cm}
{\small $^\ddagger$ Dipartimento di Matematica, Universit\`a Modena, {e-mail: {\em unguendoli@unimore.it,
cecilia.vernia@unimore.it}}}\\
\end{center}
\vskip 1truecm

\begin{abstract}
We study the response of a spin glass system with respect to the rescaling
of its interaction random variables and investigate numerically the behaviour
of the correlation functions with respect to the volume. While for a ferromagnet
the local energy correlation functions increase monotonically with the scale and,
by consequence, with respect to the volume of the system we find that in a general
spin glass model those monotonicities are violated.
\end{abstract}

\vskip 1truecm

When spins interact in a ferromagnetic system they tend to be aligned. That simple fact is reflected
in structural properties of the statistical mechanics equilibrium state called Griffiths inequalities \cite{Gr,Gr2}
or more generally GKS \cite{KS} inequalities.

The first inequality states that the pressure of a spin system
(minus the free energy times beta) does increase with the strength
of each interaction among spins. The second says that the
correlation among any set of spins increases with respect to the
strength of the interaction of any other set of them.

Since the strength of the interaction allows to switch on and off new parts of the system,
the mentioned monotonicity properties can be easily turned into new ones like
monotonicity with respect to the volume or with respect to the system dimensionality. All those
properties are at the origins of the fruitful applications of those inequalities to prove
rigorous results in statistical mechanics, like the existence of the thermodynamic limit for pressure and
correlations, bounds for critical temperatures and exponents \cite{BS} and their mutual relations for different systems.

While the physical meaning of the Griffiths inequalities was clear and well understood much earlier than rigorously proved,
for disordered systems like spin glasses there are no {\it a priori} evident monotonicity
properties due to the lack of ferromagnetism and the presence of competition (frustrated loops \cite{MPV}).

The first case to be understood has been the Gaussian
interaction. The use of the partial integration led to the first general proof of existence and
monotonicity of thermodynamic limit for the pressure of a d-dimensional spin glass model with general
potential \cite{CG}. Defining the potential
\be
\label{sgp}
U_\Lambda(J,\s) = \sum_{X\subset\Lambda} J_X \sigma_X \; ,
\ee
where the coefficients  $J_X$ are a Gaussian family distributed as
\bea
{\rm Av}(J_X) = 0 \; ,\quad
{\rm Av}(J_XJ_Y) = \Delta^2_X\delta_{X,Y}  \;
\eea
with its associated random Gibbs-Boltzmann state $\omega$ and the quenched measure as
\be
<-> \; = \; {\rm Av}\left[\omega(-)\right]
\ee
a straightforward computation \cite{CG} gives
\be\label{cl1}
<J_X\sigma_X> \; = \; \Delta^2_X {\rm Av}(1-\omega^2_X) \; ,
\ee
which turns out to be positive by inspection. Although it was clear that the former positivity
comes from convexity, it was recognized only few years later that the same result holds in full generality
for random interaction $J_X$ with zero average and not only for centered Gaussian variables:
in \cite{CL} it is proved that (\ref{cl1}) can be simply derived from thermodynamic convexity.

A natural perspective to look at the former inequalities is to consider
the deformation of the general centered random variable $J_X$ as $\lambda_X J_X$ with $\lambda_X>0$.
The quenched pressure $P$ as a function of the set of  lambda's has first and second derivatives

\be\label{i1}
\frac{\partial P}{\partial \lambda_X} \; = \; <J_X\sigma_X> \; \ge \; 0
\ee

\be\label{i2}
\frac{\partial^2 P}{\partial \lambda_X \partial \lambda_Y} \; = \;
\frac{\partial <J_X\sigma_X>}{\partial \lambda_Y} \; =
\; {\rm Av}\left[J_XJ_Y(\omega_{XY}-\omega_X\omega_Y)\right]
\ee

The two quantities (\ref{i1}) and (\ref{i2}) have been extensively studied in $d=1$ with nearest neighboor
interaction and periodic (or free) boundary
conditions in \cite{CU}: the sign of (\ref{i1}) remains positive also by shifting on positive values
the $J$ averages; the value of (\ref{i2}) turns out to be non positive for zero mean and positive variance interactions
and changes its sign when crossing a line in the mean-variance plane toward the ferromagnetic regime of
zero variance and positive mean.

The present paper deals with the study of the sign of (\ref{i2}) for higher dimensions or different
topologies for the case with zero average interaction. Our findings can be summarized as follows:
(\ref{i2}) doesn't have a definite sign. An explicit counterexample is found, for instance, in the case
of a nearest neighboor spin chain with an extra interaction connecting two distant spins.
In principle, a specific topology could not affect the monotonicity in the volume. For that reason we test
numerically the nearest neighboor correlation function for two and three dimensional systems of increasing
size and find an oscillating behaviour.

Let consider a closed chain of six spins with nearest neighboor interaction with
one added interaction between the spins $2$ and $5$. The dependence of the partition function
on the couplings $J_{1,2}$ e $J_{2,3}$ is:
\ba
Z &=& \sum_{\s} \exp \left( \b \sum J_{i,j} \s_i \s_j \right) = \\
&=& a \cosh(\b J_{1,2}) \cosh(\b J_{2,3}) +  b \sinh(\b J_{1,2}) \cosh(\b J_{2,3})  + \\
&& \qquad \qquad + c \cosh(\b J_{1,2}) \sinh(\b J_{2,3})
+ d \sinh(\b J_{1,2}) \sinh(\b J_{2,3})
\ea
where the four coefficients are:
\ba
a &=& \left( \prod_{(i,j) \neq (1,2), (2,3)} \cosh(\b J_{i,j}) \right)
\cdot \sum_{\partial B = \emptyset} \left( \prod_{(i,j) \in B} \tanh(\b J_{i,j}) \right) \\ \\
b &=& \left( \prod_{(i,j) \neq (1,2), (2,3)} \cosh(\b J_{i,j}) \right)
\cdot \sum_{\partial B = (1,2)} \left( \prod_{(i,j) \in B} \tanh(\b J_{i,j}) \right) \\ \\
c &=& \left( \prod_{(i,j) \neq (1,2), (2,3)} \cosh(\b J_{i,j}) \right)
\cdot \sum_{\partial B = (2,3)} \left( \prod_{(i,j) \in B} \tanh(\b J_{i,j}) \right) \\ \\
d &=& \left( \prod_{(i,j) \neq (1,2), (2,3)} \cosh(\b J_{i,j})
\right) \cdot \sum_{\partial B = (1,2) \cup (2,3)} \left(
\prod_{(i,j) \in B} \tanh(\b J_{i,j}) \right) \ea The truncated
correlation function is now:
\ba \omega_{12,23} - \omega_{12} \, \omega_{23}
= 16 \; \frac{ad -bc}{Z^2}
\ea
By Gauge invariance the Bernoulli
random model can be reduced to one in which the randomness is
concentrated on the two couplings $J_{1,2} = \pm 1$, $J_{2,3} = \pm
1$ with probability $1/2$,
and the remaining others $J_{i,j} =1$. \\
Let introduce the notation: $C := \cosh(\b)$, $S:= \sinh(\b)$,  $T:= \tanh(\b)$, and:
$T_{12} := \tanh(\b J_{1,2})$,
$T_{23} := \tanh(\b J_{2,3})$ (and similarly for $\sinh$ and $\cosh$).

We will indicate with $Z(+,+)$, $Z(+,-)$, $Z(-,+)$ and $Z(-,-)$ the partition functions computed with fixed
values of $J_{1,2}$ e $J_{2,3}$.  \\
\noindent An explicit computation gives:
\ba
a = C^5; \quad b = c = C^5 \, T^3; \quad d = C^5 \, T^4 \quad \Rightarrow
\ea
\ba
\Rightarrow ad -bc = C^{10} \, T^4 \, (1 - T^2)
\ea
and
\ba
Z = C^5 \, C_1 \, C_2 \, [ 1 + T_{12} \, T^3 + T_{23} \, T^3 + T_{12} \, T_{23} \, T^4 ] \quad \Rightarrow
\ea
\ba
Z(++) &=& C^7 [ 1 + 2 T^4 + T^6 ] \\
Z(+-) = Z(-+) &=& C^7 [1 - T^6] \\
Z(--) &=& C^7 [ 1 - 2 T^4 + T^6 ]
\ea
Finally:
\ba
\av{ J_{1,2}  J_{2,3} (\omega_{12,23}  -  \omega_{12} \, \omega_{23}) } = 16 \, C^{10} \, T^4 \, (1 - T^2) \,
\av{\frac{J_{1,2} \, J_{2,3}}{Z^2}} =
\ea
\ba
&& \qquad \qquad = 16 C^{10} \, T^4 \, (1 - T^2) \, \left\{ \frac{1}{Z^2(++)} + \frac{1}{Z^2(--)} -
\frac{2}{Z^2(+-)} \right\} = \\ \\
&& \qquad \qquad = \frac{16 T^4 \, (1-T^2)}{C^4}  \left\{ \frac{1}{(1 + 2 T^4 + T^6)^2} +
\frac{1}{(1 - 2 T^4 + T^6)^2} - \frac{2}{(1 - T^6)^2} \right\} = \\ \\
&& \qquad \qquad = \frac{16 T^4 \, (1-T^2)}{C^4}  \, \frac{8\, T^6 \, [-1 + 3T^2 -2T^6 +2T^8 -4T^{10} -
T^{12} + 3 T^{14}]}
{(1 + 2 T^4 + T^6)^2 \cdot (1 - 2 T^4 + T^6)^2 \cdot (1-T^6)^2}
\ea
In order to compute the sign, we notice that the square parenthesis term is:
\ba
& [-1 + 3T^2 -2T^6 +2T^8 -4T^{10} - T^{12} + 3 T^{14}] = \qquad \qquad \qquad &\\ \\
&\qquad \qquad \qquad {\ds =\left[ \frac{C^2(S^2 - 1) (C^8 - S^8) + 2 S^4 (C^8-S^8) + S^8 (C^2 + S^2)}{C^{14}} \right]} &
\ea
The only term with possible sign change is $(S^2 - 1)$. For small $\b$ the leading term is then
$- \cosh^{10}(\b)$, which gives a negative contribution, while for large $\b$ everything is positive.
\\
A plot of the function $\av{ J_{1,2}  J_{2,3} (\omega_{12,23}  -  \omega_{12} \, \omega_{23}) }$ shows a change of sign
around \linebreak $\b = 0.695$.

The numerical test is performed for $d$-dimensional cubic lattices $\Lambda$ of volume $N=L^d$, with $d=2,3$.
We analyze two cases of quenched disorder: the Bernoulli couplings with $J_{i,j}=\pm 1$
and the Gaussian couplings with zero mean and unit variance.

Given a spin configuration
$\sigma$
for a system of linear size $L$, we consider the observable:
\begin{equation}
Av[J_b\omega_b]\label{corr}
\end{equation}
where $b=(i,j)$ with $i,j\in\Lambda$, $|i-j|=1$, $\omega_b$ is the thermal average of the
quantity $\sigma_i\sigma_j$
and $Av[\cdot]$ is the average over the quenched disorder.

With a parallel-tempering algorithm \cite{HN} we investigate the
correlation (\ref{corr}) for lattice sizes ranging from $L=3$ to $L=24$ in the case $d=2$
and from $L=3$ to $L=10$ in the case $d=3$. For each size we consider at least
$2048$ disorder realizations and, in order to thermalize the large sizes, we choose
up to $37$ temperature values in the range $0.5\le t\le 2.3$, in which the critical temperatures
of the three dimensional models ($t_c\simeq 0.95$ for the Gaussian model \cite{MPRL} and
$t_c\simeq 1.15$ for the Bernoulli model \cite{BCF}) are contained.
The thermalization in the parallel tempering procedure is tested by checking the symmetry
of the probability distribution
for the standard overlap $q$ under the transformation
$q \to -q$. Moreover, for the Gaussian coupling case it is available another
thermalization test: the internal energy can be calculated
both as the temporal mean of the Hamiltonian or, using integration by parts, as the expectation of a simple
function of the link overlap \cite{Co}.
We checked that with our thermalization steps both measurements converge
to the same value.
All the parameters used in the simulations are reported in Tab.\ref{t:para}.

\begin{table}
\begin{center}
\begin{tabular}{|c|c|c|c|c|c|c|c|}\hline
\multicolumn{8}{|c|}{Two dimensional lattice $L^d, d=2$}\\ \hline
$L$ & Therm    & Equil   & Nreal  & $n_{t}$ & $\delta t$ & $t_{min}$ & $t_{max}$\\ \hline \hline
$3-12$ & $50000$  & $50000$ & $4096$ &   $19$    & $0.1 $     &  $0.5$    &   $2.3$ \\ \hline
$16$   & $50000$  & $50000$ & $2048$ &   $19$    & $0.05$     &  $0.5$    &   $2.3$ \\ \hline
$24$   & $50000$  & $50000$ & $2600$ &   $19$    & $0.05$     &  $0.5$    &   $2.3$ \\ \hline\hline
\multicolumn{8}{|c|}{Three dimensional lattice $L^d, d=3$}\\ \hline
$L$ & Therm    & Equil   & Nreal  & $n_{t}$ & $\delta t$ & $t_{min}$ & $t_{max}$\\ \hline \hline
$3-6$ & $50000$  & $50000$ & $2048$ &   $19$    & $0.1 $     &  $0.5$    &   $2.3$ \\ \hline
$8$   & $50000$  & $50000$ & $2680$ &   $19$    & $0.1 $     &  $0.5$    &   $2.3$ \\ \hline
$10$  & $70000$  & $70000$ & $2048$ &   $37$    & $0.05$     &  $0.5$    &   $2.3$ \\ \hline
\end{tabular}\caption{Parameters of the simulations: linear system size, number of
sweeps used for thermalization, number of sweeps for measurement of the observable, number
of disorder realizations, number of temperature values allowed in the parallel tempering procedure,
temperature increment, minimum and maximum temperature values.}\label{t:para}
\end{center}
\end{table}

\begin{figure}
    \setlength{\unitlength}{1cm}
          \centering
               \includegraphics[width=11cm,height=6cm]{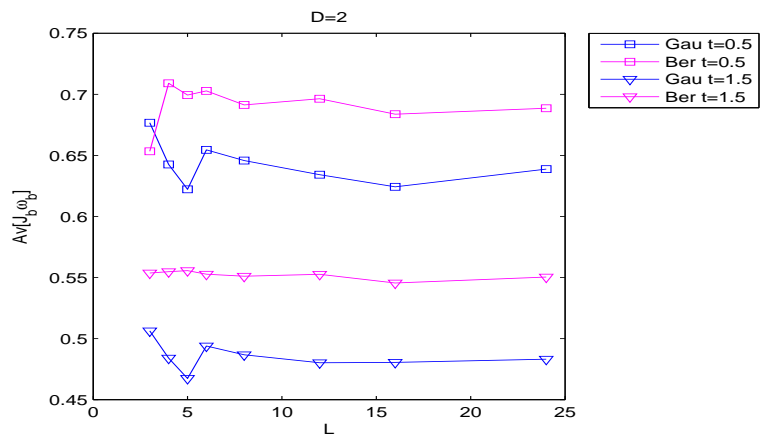}
               \caption{$Av[J_b\omega_b]$ as a function of $L$ for
               the two dimensional lattice with gaussian and Bernoulli coupling,
               for two different temperature values $t=0.5$ and $t=1.5$ each.}

\label{fig1}
\end{figure}

\begin{figure}
    \setlength{\unitlength}{1cm}
          \centering
               \includegraphics[width=11cm,height=6cm]{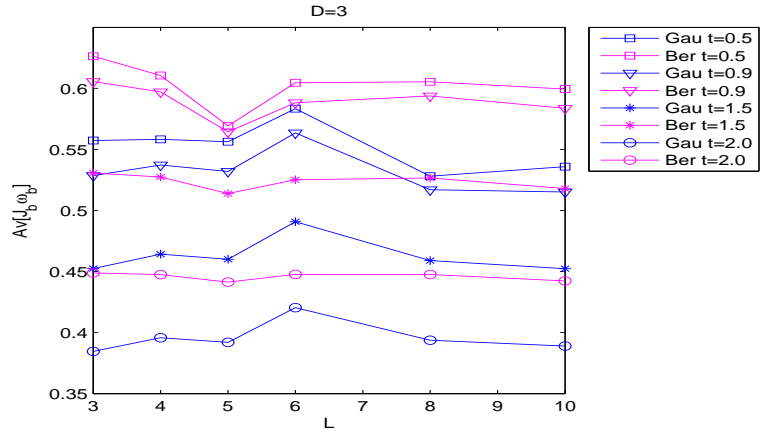}
               \caption{$Av[J_b\omega_b]$ as a function of $L$ for
               the three dimensional lattice with gaussian and Bernoulli coupling and
               for four different temperature values $t=0.5$, $t=0.9$, $t=1.5$ and $t=2$ each.}
\label{fig2}
\end{figure}

The numerical results are displayed in Fig.\ref{fig1} and in Fig.\ref{fig2}, where
the correlation (\ref{corr}) is represented as a function of the linear system size $L$
for different temperatures both for the two dimensional system and for the three dimensional
one.
We find that the correlation oscillates with respect to $L$, independently of the
couplings (Bernoulli and gaussian), of the temperatures and of the dimension $d$ ($d=2,3$).

The results presented here show that the monotonocity properties typical of ferromagnetic systems
are clearly violated for spin glass models. Further effort is necessary to establish if
the quantity (\ref{i2}) can keep a definite sign for specific lattice geometries or may
depend on the relative position of the two sets $X$ and $Y$.

{\bf Acknowledgments}. P.C. thanks Hal Tasaki for suggesting the topology used in analytical
example. Cristian Giardin\'a, Sandro Graffi, Frank Den Hollander
and Hidetoshi Nishimori are acknowledged for many useful discussions.

\begin{thebibliography}{CL}

\bibitem[CL]{CL} P.Contucci, J.Lebowitz,
{\it Correlation Inequalities for Spin Glasses},
Annales Henri Poincare, Vol. {\bf 8}, N. 8, 1461-1467, (2007)

\bibitem[CG]{CG} P.Contucci, S.Graffi,
{\it Monotonicity and thermodynamic limit},
J. Stat. Phys., Vol. {\bf 115}, Nos. 1/2, 581-589, (2004)

\bibitem[Gr]{Gr} R. B. Griffiths,
{\it Correlation in Ising Ferromagnets},
J. Math. Phys., Vol. {\bf 8}, 478-483, (1967)

\bibitem[Gr2]{Gr2} R. B. Griffiths,
{\it A proof that the free energy of a spin system is extensive},
J. Math. Phys., Vol. {\bf 5}, 1215-1222, (1964)

\bibitem[KS]{KS} D.G.Kelly, S. Sherman,
{\it General Griffiths' Inequalities on Correlations in Ising Ferromagnets},
J. Math. Phys., Vol. {\bf 9}, 466-484, (1968)

\bibitem[BS]{BS} B. Simon,
{\it The Statistical Mechanics of Lattice Gases},
Princeton University Press, (1993)

\bibitem[CU]{CU} P. Contucci, F. Unguendoli,
{\it Correlation inequalities for spin glass in one dimension},
Rend. Lincei Mat. Appl., to appear

\bibitem[MPV]{MPV} M. Mezard, G. Parisi, M. A. Virasoro,
{\it Spin Glass Theory and Beyond},
World Scientific, (1987)


\bibitem[CMN]{CMN} P.Contucci, S.Morita, H.Nishimori,
{\it Surface Terms on the Nishimori Line of the Gaussian Edwards-Anderson Model},
J. Stat. Phys., Vol. {\bf 122}, N. 2, 303-312, (2006)

\bibitem[MNC]{MNC}
S.Morita, H.Nishimori, P.Contucci,
{\it Griffiths Inequalities for the Gaussian Spin Glass},
J. Phys. A: Math. Gen., Vol. {\bf 37}, L203-L209, (2004)

\bibitem[KNA]{KNA} H.Kitatani, H.Nishimori, A.Aoki,
{\it Inequalities for the local Energy of Random Ising Models},
J. of the Physical Society of Japan, Vol. {\bf 76}, Issue 7, pp. 074711 (2007).

\bibitem[CG2]{CG2} P.Contucci, S.Graffi,
{\it On the surface pressure for the Edwards-Anderson Model},
Comm. Math. Phys. Stat. Phys., Vol. {\bf 248}, 207-220, (2004)

\bibitem[Co]{Co} P. Contucci,
{\it Replica Equivalence in the Edwards-Anderson Model},
J. Phys. A: Math. Gen., Vol. {\bf 36}, 10961-10966, (2003).

\bibitem[MPRL]{MPRL} E. Marinari, G. Parisi, and J.J. Ruiz-Lorenzo,
{\it Phase structure of the three-dimensional Edwards-Anderson spin glass},
Phys. Rev. B, Vol. {\bf 58}, 14852-14863 (1998)

\bibitem[BCF]{BCF} H.G. Ballesteros, A. Cruz, L.A. Fernandez, V. Martin-Mayor,
J. Pech, J.J. Ruiz-Lorenzo, A. Tarancon, P. Tellez, C.L. Ullod, C. Ungil,
{\it Critical Behavior of the Three-Dimensional Ising Spin Glass},
Phys. Rev. B, Vol. {\bf 62}, 14237-14245 (2000)

\bibitem[HN]{HN} K. Hukushima, K. Nemoto,
{\it Exchange Monte Carlo Method and Application to Spin Glass Simulations},
J. of the Physical Society of Japan, Vol. {\bf 65}, 1604-1608 (1996)


\end{thebibliography}
\end{document}